\documentclass[lettersize,journal]{IEEEtran}
\usepackage{amsmath,amsfonts}
\usepackage{algorithmic}
\usepackage{algorithm}
\usepackage{array}
\usepackage[caption=false,font=normalsize,labelfont=sf,textfont=sf]{subfig}
\usepackage{textcomp}
\usepackage[utf8]{inputenc}
\usepackage{stfloats}
\usepackage{booktabs}
\usepackage[caption=false,font=normalsize,labelfont=sf,textfont=sf]{subfig}
\usepackage{amssymb}
\usepackage{url}
\usepackage{bm}
\usepackage{amsmath,amssymb,amsfonts}
\usepackage{graphicx}
\usepackage{subcaption}
\usepackage[countmax]{subfloat}
\usepackage{textcomp}
\usepackage{xcolor}
\usepackage{stfloats}
\usepackage{amsmath}
\usepackage{amsmath}
\usepackage{algorithm}
\usepackage{verbatim}
\usepackage{caption}
\usepackage{graphicx}
\usepackage{cite}
\def\BibTeX{{\rm B\kern-.05em{\sc i\kern-.025em b}\kern-.08em
    T\kern-.1667em\lower.7ex\hbox{E}\kern-.125emX}}
\usepackage{amsmath, amsthm}

\begin{document}

\title{Virtual-Real Collaborated Split Learning via Model Partitioning in IRS-Assisted IoT Networks}

\author{Jiaying Di\thanks{J. Di, K. Wang and J. Xu are with the Shanghai Key Laboratory of Multidimensional
Information Processing, East China Normal University, Shanghai 200241, China, and also with the School of Communication and Electronic Engineering, East China Normal University, Shanghai 200241, China (e-mail: 51265904021@stu.ecnu.edu.cn; klwang@cee.ecnu.edu.cn; jxu@ce.ecnu.edu.cn).}, Kunlun Wang, Jing Xu, Wen Chen\thanks{W. Chen is with the Department of Electronic Engineering, Shanghai Jiao Tong University, Shanghai 200240, China (e-mail: wenchen@sjtu.edu.cn).} and Dusit Niyato\thanks{D. Niyato is with the College of Computing and Data Science, Nanyang Technological University, Singapore 639798 (e-mail: dniyato@ntu.edu.sg).}}
\maketitle
\begin{abstract}
This paper investigates a novel computation and communication co-design framework for large-scale split learning in intelligent reflecting surface (IRS)-assisted internet of things (IoT) networks integrated with digital twin (DT) technique. The considered system consists of a multi-antenna access point (AP), multiple heterogeneous user devices (UDs), and an deployed IRS to enhance both uplink and downlink transmission. The training process of a deep neural network is partitioned between devices and the AP, where a DT replica is activated to replace UDs with insufficient local computation capabilities. We formulate a delay-optimal split learning problem, which optimizes five key variables: layer partitioning points, DT assignment decisions, IRS phase shift matrix, AP downlink power allocation, and DT frequency adjustment, aiming to minimize the overall end-to-end delay under communication and computation. The proposed optimization problem is a highly coupled non-convex mixed-integer problem. Therefore, we solve using an alternating optimization approach combining closed-form updates, semidefinite relaxation (SDR), and low-complexity heuristics. Extensive simulations demonstrate that the proposed scheme significantly reduces training delay compared to conventional baselines and achieves up to 35\% delay improvement, especially under high UD density and stringent power constraints.
\end{abstract}


\section{Introduction}
The emergence of large-scale deep learning (DL) models, encompassing billions of parameters, has ushered in unprecedented opportunities for intelligent services in next-generation wireless networks\cite{9868083,10916662,10114680}. Applications such as autonomous driving, real-time video analytics, and industrial Internet of Things (IIoT) increasingly demand rapid model training and inference at the network edge\cite{Dhar2021ASO,8984361,10542407}. However, the inherent computational intensity and memory footprint of these models render traditional fully centralized training infeasible under practical wireless constraints\cite{8454520,8808168,10964638}. Specifically, end devices often suffer from limited computing capabilities, restricted battery resources, and highly variable channel conditions, resulting in a fundamental mismatch between the computational or communication demands of large models and the capabilities of edge nodes\cite{9509294,8879484,8907365,10138567}.

To alleviate the prohibitive on-device computation burden, split learning (SL) has emerged as a promising distributed training paradigm\cite{9685493,10942030}. By strategically partitioning the deep neural network (DNN) at a cut layer, SL allows the initial layers to be executed locally on the device while the remaining layers are computed at the server. This approach enables partial offloading of computation tasks, thereby reducing local processing delay and memory usage. Nevertheless, SL inevitably introduces iterative exchange of intermediate feature maps between devices and servers during both forward and backward passes. The size of these feature maps typically scales with the number of neurons and channels in early network layers, leading to substantial communication delay in bandwidth-limited wireless environments\cite{10680636,9838867}. As the number of participating devices increases, this delay rapidly becomes the dominant performance bottleneck, undermining the delay advantage of SL\cite{9721798,10041745,9562523}.

To mitigate this communication bottleneck, intelligent reflecting surfaces (IRS) have been envisioned as a transformative physical-layer technology capable of reconfiguring the wireless propagation environment in a cost- and energy-efficient manner\cite{10810732}. Comprising a large number of nearly passive reflecting elements, an IRS can dynamically adjust the phase shifts of each element to coherently combine reflected signals at the intended receiver, thereby boosting the effective channel gain and mitigating interference. When integrated into SL systems, an IRS can substantially enhance both uplink and downlink transmission rates, reducing the end-to-end delay associated with feature map exchange\cite{10000645,9531372}. Furthermore, the passive nature of IRS operation ensures scalability without incurring prohibitive energy overhead, which is particularly advantageous in dense multi-user SL scenarios\cite{10124759}.

Despite these enhancements, SL performance remains constrained by the heterogeneous computing capabilities across devices. In many realistic deployments, a subset of devices may be unable to complete their assigned model partition within the strict synchronization window dictated by the global training schedule. This performance disparity can stall the entire training process, leading to idle server resources and degraded convergence speed. To address this issue, digital twin (DT) technique acts as a useful tool\cite{9540177}. A DT is a high-fidelity virtual replica of a physical device instantiated at the access point (AP) or edge server, capable of executing the device-side computation on behalf of delay-constrained devices\cite{9540143,11088790,10941887}. By adjusting the DT's operating frequency via dynamic CPU/GPU resource allocation, the system can ensure that the completion times of substituted tasks align with those of the faster devices, thereby preserving synchronous training and minimizing idle time at the server\cite{9540131,9540135}.

\section{Related Work}

Early studies on IRS mainly focus on joint active and passive beamforming optimization. The authors in \cite{WuZhangTcom2021} proposed an alternating optimization framework that jointly designs the transmit beamforming at the AP and the phase shifts at the IRS to maximize the achievable rate. Similarly, The authors in \cite{HuangTWC2019} optimized the IRS phase shift matrix for energy-efficient communications, and the authors in \cite{YuJSAC2020} extended IRS design to robust and secure wireless transmissions. While these studies provide foundations for IRS-aided communications, they mainly consider physical-layer metrics such as signal-to-noise ratio (SNR) or secrecy capacity, without addressing the computational delay induced by large-scale deep learning models.

To complement these physical-layer advancements, another line of research targets the computation aspect through edge/cloud co-inference and split learning, which aim to alleviate on-device processing burdens by partitioning DL models between end devices and servers. For example, the authors in \cite{KangASPLOS2017} introduced "Neurosurgeon", an adaptive DNN partitioning system between mobile devices and the cloud. More recently, The authors of \cite{WuJSAC2023} proposed Cluster-based Parallel Split Learning (CPSL), which adopts a “first-parallel-then-sequential” training strategy by clustering devices for parallel training to reduce delay in large-scale SL scenarios. The authors in \cite{WuTPDS2022} introduced High-Throughput Deep Learning (HiTDL), a runtime framework for managing multiple hybrid-deployed DNNs at the edge, aiming to maximize overall throughput while meeting SLAs.
It builds performance models for delay and throughput, generates SLA-compliant partition plans, and selects optimal plans via a fairness-aware multiple-choice knapsack solution. The authors in \cite{10096914} leveraged split computing to dynamically choose the optimal partition point of a DNN between end devices and edge servers, enabling faster inference without accuracy loss under varying network and server conditions.

While IRS-based physical-layer optimization can improve communication efficiency, and split learning can reduce computational load, there remains a gap in achieving full-stack coordination between communication and computation—especially under dynamic network conditions. DT technologies have recently emerged as a promising enabler to bridge this gap by providing real-time virtual replicas of physical systems, thereby enhancing network intelligence and predictability. For example, the authors in \cite{10608284} proposed a federated DT (FDT) implementation methodology that enables the construction of large-scale DT systems by interconnecting and federating multiple single DT instances, each representing a physical object, system, or process. The proposed approach introduces dedicated management, validation, and federation mechanisms to aggregate outputs from distributed DTs into a unified large-scale model. Another work is the DT Network (DTN) architecture proposed in \cite{9540177}, which envisions a future network composed of four layers: the Physical Network, Data Lake, Digital Twin Layer, and Network Application Layer.  This architecture enables advanced functions such as fault self-healing and performance optimization, demonstrating the potential of DT to provide real-time monitoring and predictive management for complex communication systems—features that could be further enhanced when integrated with IRS-assisted communication and split learning–based computation strategies.

On the other hand, several studies focus on joint communication–computation optimization. The authors in \cite{YouTWC2017} optimized energy-efficient resource allocation for MEC systems. The authors in \cite{8488502} proposed a novel UD cooperation approach for mobile edge computing (MEC), where a helper node assists the UD in both computation and communication through a four-slot transmission protocol, jointly optimizing offloading time, transmit power, and CPU frequencies under partial and binary offloading models to minimize total energy consumption. However, these methods are not tailored for the high computational complexity and data dependency in large-model split learning, nor do they incorporate DT as a computation backup mechanism.

In summary, existing IRS works excel in channel and beamforming optimization but rarely account for the computational layer. Edge/cloud co-inference and split learning research focuses on model partitioning but neglects the impact of IRS-assisted transmission. DT-based MEC frameworks improve network adaptability but often ignore the tight coupling between computation offloading and physical-layer optimization.

\subsection{Contributions}
In this work, we address the challenges of heterogeneous device capabilities, limited wireless resources, and stringent end-to-end delay in large-scale distributed deep model training. We propose a multi-user SL framework via model partitioning and DT in IRS-assisted IoT networks. IRS enhances uplink and downlink gains to reduce communication delay, while DT creates virtual replicas for devices unable to meet delay requirements, alleviating computing bottlenecks. SL partitions the deep model into front-end and back-end sub-models for flexible, parallel computation. We formulate a joint communication–computation optimization problem to minimize overall model training delay, involving split point selection, DT decision, IRS phase-shift design, AP power allocation, and DT frequency adjustment. We solve this mixed-integer non-convex problem via a five-step alternating optimization algorithm, combining SDR, convex programming, discrete search, and heuristics, with real deep model training embedded in each iteration. Simulations with realistic channel models and PyTorch-based partitioning show that our framework outperforms Full Local Inference, Full Offloading via IRS, GA-based Hybrid Inference and ADMM-based Hybrid Inference in delay, power efficiency, and task completion, and reveal the influence of IRS size, AP power budget, and DT substitution rate on performance.

The main contributions of this work are summarized as follows:

\begin{itemize}
    \item \textbf{Unified SL--IRS--DT framework:} We propose a novel multi-user wireless edge learning framework that integrates SL, IRS, and DT technologies. We design the framework for large-scale deep model training, where IRS enhances both uplink and downlink transmission, while DT compensates for heterogeneous and insufficient computing capabilities at the device side.
    
    \item \textbf{End-to-end delay-aware joint optimization:} We formulate a mixed-integer non-convex problem that jointly optimizes the cut-layer selection for each UD, device–DT assignment decisions, IRS phase shift configuration, downlink transmit power allocation, and DT frequency adjustment. The objective is to minimize the total end-to-end delay while satisfying communication, computation, and power constraints.
    
    \item \textbf{Alternating optimization algorithm:} To tackle the challenging coupling between discrete and continuous variables, we develop a five-step alternating optimization algorithm. The algorithm employs SDR for IRS phase optimization, convex programming for power allocation, and exhaustive search for cut-layer and assignment decisions.
    
    \item \textbf{Comprehensive simulation and benchmark comparison:} We implement the proposed framework with realistic multi-antenna channel models and actual deep neural network partitioning in PyTorch. Extensive simulations demonstrate that the proposed method consistently outperforms four representative baselines (Full Local Inference, Full Offloading via IRS, GA-based Hybrid Inference, ADMM-based Hybrid Inference) in terms of delay reduction, power efficiency, and balanced computation load.
\end{itemize}

\par Notations: Throughout this paper, bold letters denote matrices. ${{\mathbb{C}}^{m\times n}}$ denotes the complex space of dimension $m\times n$. $(\cdot)^H$ trace $(\cdot)$, $||\cdot||$ and diag $\{\cdot\}$ denote conjugate-transpose, trace, Euclidean norm and diagonal operation, respectively.

\begin{table}[htbp]
\centering
\caption{\small Frequently Used Notations}
\begin{tabular}{l c}
    \toprule
    \textbf{Definition} & \textbf{Notation} \\
    \midrule
    Channel coefficient from AP to IRS & $\mathbf{G} \in \mathbb{C}^{N_{\mathrm{AP}} \times N_{\mathrm{IRS}}}$ \\
    Channel coefficient from IRS to UD $k$ & $\mathbf{h}_{\mathrm{IRS},k} \in \mathbb{C}^{N_{\mathrm{IRS}} \times 1}$ \\
    Direct channel coefficient from AP to UD $k$ & $\mathbf{h}_{\mathrm{AP},k} \in \mathbb{C}^{N_{\mathrm{AP}} \times 1}$ \\
    Receiving beamforming vector at AP for UD $k$ & $\mathbf{w}_k \in \mathbb{C}^{N_{\mathrm{AP}} \times 1}$ \\
    Transmit beamforming vector for UD $k$ & $\mathbf{v}_k \in \mathbb{C}^{N_{\mathrm{AP}} \times 1}$ \\
    Binary decision variable ($1$=local, $0$=DT) & $\alpha_k$ \\
    Computational load of layer $m$ for UD $k$ & $C_k^{(m)}$ \\
    Local CPU frequency of UD $k$ & $f_k^{\mathrm{loc}}$ \\
    Additional AP frequency of DT for UD $k$ & $\Delta f_k$ \\
    Downlink transmit power allocated to UD $k$ & $P_{\mathrm{AP},k}$ \\
   
    Total AP transmit power budget & $P_{\mathrm{total}}$ \\
     Split layer index for UD $k$ & $m_k$ \\
    Phase shift matrix of IRS & $\mathbf{\Theta} $ \\
    Noise power & $\sigma^2$ \\
    Weight factor in objective function & $\lambda$ \\
    \bottomrule
\end{tabular}
\label{tab:notation}
\end{table}

\section{System Model and Network Architecture}
In this work, we consider a novel IRS-assisted SL IoT networks consisting of $\mathit{K}$ user devices (UDs), an IRS, a multi-antenna AP. The overall network structure is illustrated in Fig.1, which consists of the following key components. There are a set of $K$ wireless UDs, denoted as $\mathcal{K} = \{1, 2, \ldots, K\}$, each participating in the training of a large-scale deep learning model. Each UD is equipped with limited computational resources and is responsible for training a partial segment of the global model. There is also a centralized AP equipped with $N_{\text{AP}}$ antennas, which performs model aggregation and completes the server-side part of the model training. An IRS composed of $N_{\text{IRS}}$ passive reflecting elements is deployed between the UDs and the AP. The IRS reflects signals from both uplink and downlink directions to enhance the quality of communication channels. The communication between UDs and the AP is realized exclusively via a passive IRS, which is strategically deployed to enhance transmission through programmable phase shifts. The AP is equipped with multiple antennas to support simultaneous multi-user communication, while the IRS, reflecting elements, acts as a passive relay to compensate for the absence of direct links. Each UD has a unique computation profile and channel state, necessitating individualized optimization for model split point selection, communication strategy, and DT offloading. For UDs whose computation capacity is insufficient to process the assigned model layers, a virtual DT instance is instantiated at the AP to perform surrogate computation. In our work, the AP is also equipped with edge computing capabilities and thus functions as a MEC server. This design is consistent with recent works where AP and MEC server are often regarded interchangeably as a co-located entity, since modern access points are typically embedded with computation resources\cite{9027954}. Such an integration eliminates the additional backhaul delay between the AP and a separate edge server, thereby ensuring low-latency task execution and making the model more tractable. Each DT is allocated an adjustable computation frequency to meet delay constraints. Each UD participates in split learning by training the first $m_k$ layers of a global model locally, while the remaining $M - m_k$ layers are processed at the AP. Let $\mathcal{M} = \{1, 2, \ldots, M\}$ denote the set of model layers, and $m_k \in \mathcal{M}$ be the cut layer selected by UD $k$.
The system is designed to collaboratively train a large-scale deep neural network under stringent delay constraints and heterogeneous computational capabilities across UDs. Specifically, the global model is divided into two segments: the front part is computed locally by each UD, while the remaining part is processed at the AP or cloud. To handle the limitations of user-side computation, we introduce a DT mechanism, allowing the system to dynamically offload intractable local computation to virtual DT agents hosted at the server.

\begin{figure}[t]
    \centering
    \includegraphics[width=0.48\textwidth]{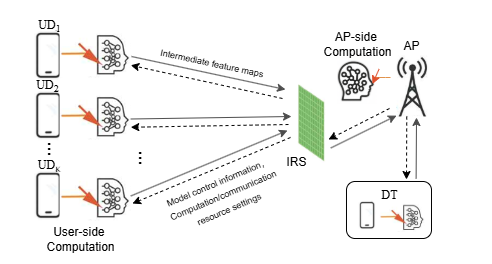}
    \caption{An edge intelligence system integrating IRS-assisted communication, split learning, and digital twin computation backup.}
    \label{fig:delay_vs_k}
\end{figure}
\subsection{Network Model}

Each UD holds private training data and collaboratively participates in the training of a global deep neural network model with $M$ layers. For each UD $k$, the model is split at layer $m_k$, meaning the first $m_k$ layers are computed locally and the remaining $M - m_k$ layers are computed at the AP. The value of $m_k$ is dynamically optimized. To support flexible and reliable communication, especially in scenarios where direct AP-to-UD links are blocked or weak, an IRS is deployed to assist both uplink and downlink signal propagation\cite{9685493,10942030}.

The communication process involves the following four transmission phases:

\begin{enumerate}
    \item \textbf{Downlink Phase 1 (AP $\rightarrow$ IRS)}: 
    The AP broadcasts input data (e.g., global model parameters, labels, or split-layer weights) to the UDs via the IRS. The transmitted signal is given by:
    \begin{equation}
        \mathbf{x}_{\mathrm{AP}} = \sum_{k=1}^{K} \mathbf{w}_k s_k,
    \end{equation}
    where $\mathbf{w}_k \in \mathbb{C}^{N_{\mathrm{AP}}}$ is the downlink beamforming vector for UD $k$, and $s_k$ is the information-bearing signal, such as training data batches or global model parameters. It is assumed $\mathbb{E}[|s_k|^2] = 1$.

    \item \textbf{Downlink Phase 2 (IRS $\rightarrow$ UD)}: 
    The IRS reflects the downlink signal toward each UD. The IRS applies a diagonal phase-shift matrix
    \begin{equation}
        \boldsymbol{\Theta} = \mathrm{diag}(e^{j\theta_1}, \dots, e^{j\theta_{N_{\mathrm{IRS}}}}),
    \end{equation}
    and the effective channel to UD $k$ becomes
    \begin{equation}
        \mathbf{h}_k^{\mathrm{eff}} = {\mathbf{h}_k^{\mathrm{IRS}}}^{H}  \boldsymbol{\Theta} \mathbf{G},
    \end{equation}
    where $\mathbf{G} \in \mathbb{C}^{N_{\mathrm{IRS}} \times N_{\mathrm{AP}}}$ is the AP-to-IRS channel and $\mathbf{h}_k^{\mathrm{IRS}} \in \mathbb{C}^{N_{\mathrm{IRS}}}$ is the IRS-to-UD-$k$ channel.

    \item \textbf{Uplink Phase 1 (UD $\rightarrow$ IRS)}:
    After computing the first $m_k$ layers locally, UD $k$ obtains an intermediate feature vector $\mathbf{z}_k$ (i.e., the output of layer $m_k$). This output is then transmitted to the AP for subsequent processing. The UD sends
    \begin{equation}
        \mathbf{x}_k^{\mathrm{UL}} = \mathbf{v}_k \mathbf{z}_k,
    \end{equation}
    where $\mathbf{v}_k$ is the uplink transmit beamforming vector (e.g., scalar if single-antenna UD). The IRS is on the transmission path and reflects the signal along the way.

    \item \textbf{Uplink Phase 2 (IRS $\rightarrow$ AP)}: 
    The IRS reflects the uplink signals from each UD toward the AP. The AP receives
    \begin{equation}
        y_k^{\mathrm{AP}} = \mathbf{g}_k^H \boldsymbol{\Theta} \mathbf{h}_k^{\mathrm{IRS}} \mathbf{x}_k^{\mathrm{UL}} + n_k,
    \end{equation}
    where $\mathbf{g}_k \in \mathbb{C}^{N_{\mathrm{AP}} \times N_{\mathrm{IRS}}}$ is the IRS-to-AP channel for UD $k$ and $n_k$ is the additive white Gaussian noise (AWGN) at the AP.
\end{enumerate}

\subsection{Digital Twin Model}

The DT is a virtual replica of a physical UD device deployed at the AP. When a UD $k$ lacks sufficient local computational resources to execute the allocated model layers, a virtual twin of that UD is instantiated in the cloud. This ensures reliable and timely execution of the split learning process. This virtual instance emulates the local computation and generates intermediate features required for the remainder of the model inference pipeline.

In our model, whether UD $k$ has a virtual copy or not is determined by a binary variable $\alpha_k \in \{0,1\}$, where $\alpha_k = 1$ means UD $k$ computes locally and $\alpha_k = 0$ means virtual clone of UD $k$ at AP performs the computation.

Let $C_k^{(m_k)}$ denote the total computational task of the first $m_k$ layers for UD $k$, and $f_k^{\mathrm{loc}}$ be the UD's local CPU frequency. The local computation delay is obtained as follows:
\begin{equation}
    T_k^{\mathrm{loc}} = \frac{C_k^{(m_k)}}{f_k^{\mathrm{loc}}}.
\end{equation}

If the computational task is offloaded to the DT, the local CPU frequency is adjusted by a tunable parameter $\Delta f_k$ to ensures that rapid response, leading to
\begin{equation}
    T_k^{\mathrm{DT}} = \frac{C_k^{(m_k)}}{f_k^{\mathrm{loc}} + \Delta f_k}.
\end{equation}

The final execution delay of UD $k$ becomes
\begin{equation}
    T_k = \alpha_k T_k^{\mathrm{loc}} + (1 - \alpha_k) T_k^{\mathrm{DT}}.
\end{equation}

The DT strategy is jointly optimized with the cut layer $m_k$, power allocation $\mathbf{P}$, IRS phase shift matrix $\boldsymbol{\Theta}$, and frequency offset $\Delta f_k$ to minimize the overall system delay.

\subsection{Delay Model}

In the considered split learning system, the total delay for UD $k$ comprises the following components: downlink communication delay $T_k^{\mathrm{DL}}$, uplink communication delay $T_k^{\mathrm{UL}}$, and computation delay $T_k$, which may occur either locally or on the DT. The overall delay for UD $k$ is given by
\begin{equation}
    T_k^{\mathrm{total}} = T_k^{\mathrm{DL}} + T_k^{\mathrm{UL}} + T_k.
\end{equation}

\paragraph{Downlink Communication Delay $T_k^{\mathrm{DL}}$}  
The AP transmits data (such as model parameters or labels) to UD $k$ through the IRS. 

Let power $P_{\mathrm{AP},k}$ and B denotes the transmit power allocated by the AP for downlink transmission to UD $k$ and bandwidth, respectively. Then, the downlink data rate is given by
\begin{equation}
    R_k^{\mathrm{DL}} = B \log_2 \left(1 + \frac{P_{\mathrm{AP},k} |\mathbf{h}_k^{\mathrm{eff}}|^2}{\sigma^2} \right),
\end{equation}
where $\sigma^2$ represents the noise power. Let $D_k^{DL}$ denote the transmitted data size, then the downlink delay is given by
\begin{equation}
    T_k^{\mathrm{DL}} = \frac{D_k^{\mathrm{DL}}}{R_k^{\mathrm{DL}}}.
\end{equation}

\paragraph{Uplink Communication Delay $T_k^{\mathrm{UL}}$}  
After local computation of the first $m_k$ layers, UD $k$ transmits intermediate result $\mathbf{z}_k$ to the AP via the IRS. The effective uplink channel is
\begin{equation}
    \mathbf{g}_k^{\mathrm{eff}} = \mathbf{g}_k^H \boldsymbol{\Theta} \mathbf{h}_k^{\mathrm{IRS}},
\end{equation}
where $\mathbf{g}_k \in \mathbb{C}^{N_{\mathrm{AP}} \times N_{\mathrm{IRS}}}$ represents the channel of the link the IRS-to-AP. Given UD power $P_k^{\mathrm{UL}}$ = $\left |w_{k}   \right | ^{2} $, the uplink data rate is given by
\begin{equation}
    R_k^{\mathrm{UL}} = B \log_2 \left(1 + \frac{P_k^{\mathrm{UL}} |\mathbf{g}_k^{\mathrm{eff}}|^2}{\sigma^2} \right).
\end{equation}
Given data size $D_k^{\mathrm{UL}}$, the uplink delay is
\begin{equation}
    T_k^{\mathrm{UL}} = \frac{D_k^{\mathrm{UL}}}{R_k^{\mathrm{UL}}}.
\end{equation}

\paragraph{Computation Delay $T_k$}  
The local or DT computation delay is given by
\begin{equation}
    T_k = \alpha_k \frac{C_k^{(m_k)}}{f_k^{\mathrm{loc}}} + (1 - \alpha_k) \frac{C_k^{(m_k)}}{f_k^{\mathrm{loc}} + \Delta f_k},
\end{equation}
where $\alpha_k \in \{0,1\}$ is the DT decision variable, $C_k^{(m_k)}$ is the computation cost of the first $m_k$ layers, $f_k^{\mathrm{loc}}$ is the local CPU frequency, and $\Delta f_k$ is the frequency offset for DT acceleration.

\subsection{Model Training Objective and Problem Formulation}

To ensure the accuracy of the global model during the distributed split learning process, we adopt the standard \textit{cross-entropy loss}\cite{9028113} as the training objective. For a given UD $k$ with ground-truth label vector ${y}_k$ and predicted probability vector $\hat{{y}}_k$ at the output of the neural network, the loss function is defined as
\begin{equation}
    \mathcal{L}_k = - \sum_{c=1}^{C} y_{k,c} \log \hat{y}_{k,c},
\end{equation}
where $C$ is the number of classes, $y_{k,c} \in \{0, 1\}$ is the true label indicator for class $c$, and $\hat{y}_{k,c} \in [0, 1]$ is the softmax output corresponding to class $c$. The global loss across all participating UDs is averaged as
\begin{equation}
    \mathcal{L}_{\mathrm{train}} = \frac{1}{K} \sum_{k=1}^{K} \mathcal{L}_k.
\end{equation}

Our objective is to jointly optimize the model split variable $\{m_k\}$, the offloading decision $\{\alpha_k\}$, the AP power allocation $\{P_{\mathrm{AP},k}\}$, the IRS phase-shift matrix $\boldsymbol{\Theta}$, and the DT frequency adjustment $\{\Delta f_k\}$, for minimizing the total system delay and the training loss. The optimization problem is formulated as:

\begin{align}
\mathop{\mathrm{minimize}}_{\substack{\{m_k\}, \{\alpha_k\}, \\ \{P_{\mathrm{AP},k}\}, \boldsymbol{\Theta}, \{\Delta f_k\}}} \quad & \sum_{k=1}^K T_k^{total} + \lambda \cdot \mathcal{L}_{\text{train}} \label{eq:final_obj} \\
\text{s.t.} \quad 
&  m_k \in \{1,2,\dots,M\}, \quad \forall k, \\
&  \alpha_k \in \{0,1\}, \quad \forall k, \\
&  \sum_{k=1}^{K} P_{\mathrm{AP},k} \leq P_{\text{total}}, \\
&  0 \leq P_{\mathrm{AP},k} \leq P_{\max}, \quad \forall k, \\
&  0 \leq \Delta f_k \leq \Delta f_{\max}, \quad \forall k, \\
& \theta_n \in [0, 2\pi), \quad \forall n = 1, \dots, N_{\mathrm{IRS}}, \\
& R_k^{\mathrm{UL}} \geq R_{\min}, \quad \forall k, \\
& T_k \leq T_{\max}, \quad \forall k, \\
& |\Theta_{n,n}| = 1, \quad \forall n.
\end{align}
Constraint (19) ensures that the cut layer $m_k$ is an integer between $1$ and $M$; Constraint (20) defines the binary offloading decision: $\alpha_k = 1$ means local computation, $\alpha_k = 0$ means DT-based computation. Constraint (21) ensures that the total power used by the AP does not exceed the maximum budget $P_{\text{total}}$. Constraint (22) restricts each UD's allocated transmit power within practical bounds. Constraint (23) bounds the DT frequency adjustment to ensure hardware feasibility. Constraint (24) ensures that each IRS element applies a valid phase shift in $[0, 2\pi)$. Constraint (25) guarantees that the uplink rate $R_k^{\mathrm{UL}}$ for each UD satisfies a minimum rate requirement. Constraint (26) ensures that the total delay for each UD is within a threshold $T_{\max}$.
Constraint (27) imposes unit-modulus conditions on the IRS reflection coefficients.

\section{Delay Minimization Schemes}
Given the mixed-integer and non-convex nature of the problem, we decompose the original problem into five tightly coupled subproblems. First, split layer selection determines the optimal partition of the neural network, balancing local computation and transmission overhead. Second, DT activation decision identifies whether each user executes the assigned layers locally or delegates them to a digital twin, depending on the device’s computation capability. Third, DT frequency offset calibrates the operating frequency of digital twins to provide sufficient computational redundancy and align their execution speed with system latency requirements. Fourth, IRS phase shift design optimizes the reflecting elements to maximize effective channel gain, thereby improving both uplink and downlink communication efficiency. Finally, AP power allocation distributes the transmission power budget among users to strike a balance between spectral efficiency, energy consumption, and fairness.
To tackle these subproblems, we adopt an alternating optimization strategy, iteratively updating each variable while keeping the others fixed until convergence is achieved. We separate the original problem into these five steps because each subproblem involves different mathematical structures (discrete decisions, continuous non-convex optimization, or convex resource allocation), and each has distinct physical interpretations in the system. This decomposition not only reduces the overall computational complexity but also allows us to exploit specialized optimization methods for each subproblem. Moreover, the alternating optimization strategy ensures that the coupled variables can be coordinated effectively, leading to a convergent and practically implementable solution.

\subsection{Split Layer Selection}

In this step, we optimize the split point $m_k$ for each UD $k$, which determines how many layers of the deep neural network are executed locally on the UD's device and how many are executed at the AP. The split point plays a pivotal role in affecting both the computation and communication delays as well as the overall training performance.
The total number of layers in the global model is $M$. For each UD $k$, we define the cut layer as $m_k \in \{1, 2, \ldots, M\}$, where the first $m_k$ layers are computed locally (or by the DT) and the remaining $M - m_k$ layers are computed at the AP.
Let $T_k(m_k)$ denote the total delay associated with UD $k$ when using split point $m_k$, and $L_k(m_k)$ be the training loss incurred at this split point.

The optimization objective for UD $k$ is to minimize the joint delay-loss metric, defined as follows:
\begin{equation}
    J_k(m_k) = T_k^{total}(m_k) + \lambda \cdot \mathcal{L}_{\text{train}}(m_k),
\end{equation}
where $\lambda > 0$ is a trade-off parameter between delay and learning accuracy.

We adopt an exhaustive search to determine the optimal split point for each UD. Unlike exhaustive enumeration across all users’ cut-layer combinations with exponential complexity, our approach adopts a per-user exhaustive search, which can also be regarded as a lightweight heuristic, since it linearly searches within a small candidate set for each user rather than exploring the entire global space. As $M$ is typically small, this step is computationally negligible compared to IRS optimization or power allocation, and thus practically efficient and scalable. The update process follows a simple thresholding logic as
illustrated in the following pseudocode:

\begin{algorithm}[H]
\caption{Exhaustive Search of Split Point $m_k$}
\begin{algorithmic}[1]
\FOR{each UD $k = 1$ to $K$}
    \STATE Initialize $J_k^{\mathrm{min}} \gets \infty$
    \FOR{each candidate split point $m_k \in \{1, 2, \ldots, M\}$}
        \STATE Compute local delay $T_k^{\mathrm{loc}} = \frac{C_k^{(m_k)}}{f_k^{\mathrm{loc}}}$
        \STATE Compute DT delay $T_k^{\mathrm{DT}} = \frac{C_k^{(m_k)}}{f_k^{\mathrm{loc}} + \Delta f_k}$
        \STATE Determine $\alpha_k$
        \STATE Compute total delay $T_k^{total}(m_k)$
        \STATE Estimate training loss $L_k(m_k)$
        \STATE Compute objective $J_k(m_k) = T_k^{total}(m_k) + \lambda \cdot \mathcal{L}_{\text{train}}(m_k)$
        \IF{$J_k(m_k) < J_k^{\mathrm{min}}$}
            \STATE Update $m_k^* \gets m_k$
            \STATE Update $J_k^{\mathrm{min}} \gets J_k(m_k)$
        \ENDIF
    \ENDFOR
    \STATE Set final split point $m_k \gets m_k^*$
\ENDFOR
\end{algorithmic}
\end{algorithm}

\subsection{Digital Twin Activation Decision}

After determining the optimal model split point \(m_k\) for each UD, the system evaluates whether the UD \(k\) has sufficient computational capacity to execute the first \(m_k\) layers locally. This decision is represented by the binary variable \(\alpha_k \in \{0, 1\}\).
To ensure timely execution, we adopt the following decision rule:
\begin{equation}
\alpha_k = 
\begin{cases}
1, & \text{if } f_k^{\mathrm{loc}} \geq \frac{C_k^{(m_k)}}{T_{\max}^{(k)}}, \\
0, & \text{otherwise},
\end{cases}
\label{eq:alpha_decision}
\end{equation}
where \(C_k^{(m_k)}\) denotes the computation task of the first \(m_k\) layers, \(f_k^{\mathrm{loc}}\) is the local CPU frequency of UD $k$, and \(T_{\max}^{(k)}\) represents a threshold delay derived from system delay constraints or previous iteration benchmarks.

This decision is updated in each iteration of the alternating optimization framework, as both \(m_k\) and system delay evolve. The update process follows a simple thresholding logic as illustrated in the following pseudocode:

\begin{algorithm}[H]
\caption{DT Activation Decision}
\begin{algorithmic}[1]
\FOR{each UD \(k = 1\) to \(K\)}
    \STATE Compute local delay: \(T_k^{\mathrm{loc}} = \frac{C_k^{(m_k)}}{f_k^{\mathrm{loc}}}\)
    \IF{\(T_k^{\mathrm{loc}} \leq T_{\max}^{(k)}\)}
        \STATE \(\alpha_k \gets 1\) \COMMENT{Local computation is feasible}
    \ELSE
        \STATE \(\alpha_k \gets 0\) \COMMENT{Activate DT for this UD}
    \ENDIF
\ENDFOR
\end{algorithmic}
\end{algorithm}

\subsection{Optimization of DT Frequency Offset \(\Delta f_k\)}

When a UD $k$ is determined to offload its computation to the DT, the total local computation is handled by the AP. To accelerate the DT’s response and reduce delay, the CPU frequency is augmented by a controllable offset \(\Delta f_k\), forming an adjustable computing capacity of \(f_k^{\mathrm{loc}} + \Delta f_k\).

The corresponding DT computation delay is
\begin{equation}
    T_k^{\mathrm{DT}} = \frac{C_k^{(m_k)}}{f_k^{\mathrm{loc}} + \Delta f_k},
\end{equation}
which decreases as with \(\Delta f_k\) increases. However, to avoid excessive resource usage at the AP, we impose a global constraint on the total available AP frequency budget as follows:
\begin{equation}
    \sum_{k=1}^{K} (1 - \alpha_k) \Delta f_k \leq \Delta f_{\max}, \quad \Delta f_k \geq 0.
    \label{eq:dt_freq_constraint}
\end{equation}
The objective is to minimize the total DT-related computation delay under the global AP frequency budget given as follows:
\begin{equation}
\begin{aligned}
    \min_{\{\Delta f_k\}} \quad & \sum_{k=1}^{K} (1 - \alpha_k) \cdot \frac{C_k^{(m_k)}}{f_k^{\mathrm{loc}} + \Delta f_k} \\
    \text{s.t.} \quad & \sum_{k=1}^{K} (1 - \alpha_k) \cdot \Delta f_k \leq \Delta f_{\max}, \\
                      & \Delta f_k \geq 0, \quad \forall k.
\end{aligned}
\label{eq:deltaf_opt_problem}
\end{equation}

This is a convex optimization problem and can be solved analytically using the Lagrangian method. We define the Lagrangian function as follows:
\begin{equation}
\begin{aligned}
     \mathcal{L} = &\sum_{k=1}^{K} (1 - \alpha_k) \cdot \frac{C_k^{(m_k)}}{f_k^{\mathrm{loc}} + \Delta f_k}
    + \\
    &\lambda \left( \sum_{k=1}^{K} (1 - \alpha_k) \cdot \Delta f_k -
      \Delta f_{\max} \right) 
     - \sum_{k=1}^{K} \mu_k \cdot \Delta f_k,
\end{aligned}
\end{equation}
where \(\lambda \geq 0\) is the Lagrange multiplier for the CPU budget constraint, and \(\mu_k \geq 0\) enforces non-negativity on each \(\Delta f_k\).

We take the derivative of the Lagrangian with respect to \(\Delta f_k\) and apply the KKT stationarity condition as follows:
\begin{equation}
    \frac{\partial \mathcal{L}}{\partial \Delta f_k} = 
    - (1 - \alpha_k) \cdot \frac{C_k^{(m_k)}}{(f_k^{\mathrm{loc}} + \Delta f_k)^2}
    + \lambda (1 - \alpha_k) - \mu_k = 0.
    \label{eq:kkt_derivative}
\end{equation}

For UDs with \(\alpha_k = 1\), we fix \(\Delta f_k = 0\); for \(\alpha_k = 0\), the above becomes
\begin{equation}
    \Delta f_k^* = \sqrt{ \frac{C_k^{(m_k)}}{\lambda} } - f_k^{\mathrm{loc}}.
\end{equation}

Combining with the constraint \(\Delta f_k^* \geq 0\), the closed-form solution becomes
\begin{equation}
    \Delta f_k^* = 
    \begin{cases}
    \max\left\{ \sqrt{ \frac{C_k^{(m_k)}}{\lambda} } - f_k^{\mathrm{loc}}, 0 \right\}, & \text{if } \alpha_k = 0, \\
    0, & \text{if } \alpha_k = 1.
    \end{cases}
    \label{eq:deltaf_final}
\end{equation}
The Lagrange multiplier \(\lambda\) can be updated via a bisection search to satisfy the total CPU constraint
\begin{equation}
    \sum_{k=1}^{K} (1 - \alpha_k) \Delta f_k^* = \Delta f_{\max}.
\end{equation}
The update process follows a thresholding logic as
illustrated in the following pseudocode:
\begin{algorithm}[H]
\caption{Optimization of DT Frequency Offset \(\Delta f_k\)}
\begin{algorithmic}[1]
\STATE Initialize: \(\lambda_{\min}, \lambda_{\max}, \epsilon\)
\WHILE{\(|\sum_k (1 - \alpha_k)\Delta f_k^* - \Delta f_{\max}| > \epsilon\)}
    \STATE \(\lambda \gets (\lambda_{\min} + \lambda_{\max})/2\)
    \FOR{each UD \(k = 1\) to \(K\)}
        \IF{\(\alpha_k = 0\)}
            \STATE \(\Delta f_k^* \gets \max\left\{ \sqrt{ \frac{C_k^{(m_k)}}{\lambda} } - f_k^{\mathrm{loc}}, 0 \right\}\)
        \ELSE
            \STATE \(\Delta f_k^* \gets 0\)
        \ENDIF
    \ENDFOR
    \IF{\(\sum_k (1 - \alpha_k)\Delta f_k^* > \Delta f_{\max}\)}
        \STATE \(\lambda_{\min} \gets \lambda\)
    \ELSE
        \STATE \(\lambda_{\max} \gets \lambda\)
    \ENDIF
\ENDWHILE
\end{algorithmic}
\end{algorithm}

\subsection{Optimization of IRS Phase Shift Matrix \(\boldsymbol{\Theta}\)}

The IRS phase shift matrix \(\boldsymbol{\Theta} = \mathrm{diag}(e^{j\theta_1}, \dots, e^{j\theta_{N_{\mathrm{IRS}}}})\) plays a critical role in shaping both uplink and downlink signal propagation. By appropriately configuring \(\boldsymbol{\Theta}\), the effective end-to-end channel gain in both links can be enhanced, improving data transmission rate and reducing communication delay.

We jointly consider the uplink and downlink signal strength received by the AP and UD devices, respectively. Let \(\mathbf{G} \in \mathbb{C}^{N_{\mathrm{IRS}} \times N_{\mathrm{AP}}}\) denote the channel of link AP-to-IRS. Let \(\mathbf{h}_k^{\mathrm{IRS}} \in \mathbb{C}^{N_{\mathrm{IRS}}}\) denote the channel of link IRS-to-UD$k$. Let \(\mathbf{g}_k \in \mathbb{C}^{N_{\mathrm{AP}} \times N_{\mathrm{IRS}}}\) denote the channel of link IRS-to-AP for UD $k$ (can be \(\mathbf{G}^H\) if reciprocal). The effective downlink channel for UD $k$ is
\begin{equation}
    \mathbf{h}_k^{\mathrm{DL}} = {\mathbf{h}_k^{\mathrm{IRS}}}^{H} \boldsymbol{\Theta} \mathbf{G},
\end{equation}
and the effective uplink channel is:
\begin{equation}
    \mathbf{h}_k^{\mathrm{UL}} = \mathbf{g}_k^H \boldsymbol{\Theta} \mathbf{h}_k^{\mathrm{IRS}}.
\end{equation}
Our goal is to jointly enhance the overall signal strength across all UDs by maximizing the total squared effective channel gain across uplink and downlink expressed as follows:
\begin{equation}
\begin{aligned}
    \max_{\boldsymbol{\Theta}} \quad & \sum_{k=1}^K \left( \left\| \mathbf{h}_k^{\mathrm{DL}} \right\|^2 + \left| \mathbf{h}_k^{\mathrm{UL}} \right|^2 \right) \\
    = \max_{\boldsymbol{\Theta}} \quad & \sum_{k=1}^K \left( \left\| {\mathbf{h}_k^{\mathrm{IRS}}}^H \boldsymbol{\Theta} \mathbf{G} \right\|^2 + \left| \mathbf{g}_k^H \boldsymbol{\Theta} \mathbf{h}_k^{\mathrm{IRS}} \right|^2 \right)
\end{aligned}
\label{eq:irs_objective}
\end{equation}
By defining an auxiliary variable:
\begin{equation}
    \mathbf{v} = \begin{bmatrix}
        e^{j\theta_1} \\
        \vdots \\
        e^{j\theta_{N_{\mathrm{IRS}}}}
    \end{bmatrix}, \quad |v_i| = 1.
\end{equation}

We can rewrite the objective as:
\begin{equation}
    \max_{\mathbf{v}} \quad \sum_{k=1}^{K} \left( \mathbf{v}^H \mathbf{Q}_k \mathbf{v} \right), \quad \text{s.t. } |v_i| = 1 \ \forall i,
\end{equation}
where \(\mathbf{Q}_k\) is the Hermitian matrix derived from the channel terms for UD \(k\).

This problem is non-convex due to the unit modulus constraint on \(\mathbf{v}\). We apply SDR to relax the constraint. Let \(\mathbf{V} = \mathbf{v} \mathbf{v}^H\), then the problem becomes
\begin{equation}
\begin{aligned}
    \max_{\mathbf{V}} \quad & \sum_{k=1}^{K} \mathrm{Tr}(\mathbf{Q}_k \mathbf{V}) \\
    \text{s.t.} \quad & \mathbf{V} \succeq 0, \\
                      & [\mathbf{V}]_{i,i} = 1, \quad \forall i.
\end{aligned}
\end{equation}

This is a standard convex Semidefinite Program (SDP), which can be solved using CVX or SCS.

After solving for \(\mathbf{V}^*\), we extract a feasible rank-1 solution \(\mathbf{v}\) by Gaussian randomization:
\begin{enumerate}
    \item Perform eigen-decomposition: \(\mathbf{V}^* = \mathbf{U} \mathbf{\Lambda} \mathbf{U}^H\)
    \item Sample: \(\tilde{\mathbf{v}} = \mathbf{U} \mathbf{\Lambda}^{1/2} \mathbf{z}\), where \(\mathbf{z} \sim \mathcal{CN}(0, \mathbf{I})\)
    \item Normalize: \(v_i = \frac{\tilde{v}_i}{|\tilde{v}_i|}\)
\end{enumerate}

Finally, construct the IRS phase shift matrix
\begin{equation}
    \boldsymbol{\Theta} = \mathrm{diag}(v_1, \dots, v_{N_{\mathrm{IRS}}}).
\end{equation}
The update process follows a thresholding logic as illustrated in the following pseudocode:
\vspace{1em}
\begin{algorithm}[H]
\caption{IRS Phase Shift Optimization via SDR}
\begin{algorithmic}[1]
\STATE Compute channel matrices \(\mathbf{Q}_k\) for all UDs.
\STATE Formulate \(\mathbf{V} \succeq 0\), \([\mathbf{V}]_{i,i} = 1\).
\STATE Solve SDP: \(\max \sum_k \mathrm{Tr}(\mathbf{Q}_k \mathbf{V})\)
\STATE Decompose \(\mathbf{V}^* = \mathbf{U} \mathbf{\Lambda} \mathbf{U}^H\)
\FOR{trial = 1 to \(N_{\mathrm{rand}}\)}
    \STATE Sample \(\mathbf{z} \sim \mathcal{CN}(0, \mathbf{I})\)
    \STATE \(\tilde{\mathbf{v}} \gets \mathbf{U} \mathbf{\Lambda}^{1/2} \mathbf{z}\)
    \STATE Normalize: \(v_i = \frac{\tilde{v}_i}{|\tilde{v}_i|}\)
    \STATE Compute objective with current \(\mathbf{v}\)
\ENDFOR
\STATE Select best \(\mathbf{v}\), construct \(\boldsymbol{\Theta} = \mathrm{diag}(\mathbf{v})\)
\end{algorithmic}
\end{algorithm}

\subsection{Optimization for Downlink Power Allocation}
When other variables are fixed, the downlink rate for user $k$ with power $P_{AP,k}$ is given by
\begin{equation}
R_{\text{DL},k} \;=\; B \log_2\!\Big(1 + a_k P_{AP,k}\Big), \qquad 
a_k \triangleq \frac{|h_{\text{eff},k}|^2}{\sigma^2}.
\end{equation}
The corresponding downlink transmission delay is
\begin{equation}
T_{\text{DL},k}(P_{AP,k}) \;=\; \frac{D_{\text{DL},k}}{R_{\text{DL},k}}
= \frac{D_{\text{DL},k}\ln 2}{B \,\ln(1 + a_k P_{AP,k})}.
\end{equation}
Define $c_k \triangleq \dfrac{D_{\text{DL},k}\ln 2}{B}>0$, so the system's total downlink delay minimization problem can be written as
\begin{align}
\min_{\{P_{AP,k}\}} \quad & \sum_{k=1}^{K} \frac{c_k}{\ln(1 + a_k P_{AP,k})} \label{prob:pap-obj}\\
\text{s.t.}\quad 
& \sum_{k=1}^{K} P_{AP,k} \;\le\; P_{\text{total}}, \label{cons:sum}\\
& 0 \;\le\; P_{AP,k} \;\le\; P_{\max}, \quad \forall k. \label{cons:box}
\end{align}
The objective function is a decreasing convex function with respect to $P_{AP,k}$, and the constraints are affine and box constraints, so the problem is convex and has a unique global optimum.
To handle the constraints in \eqref{cons:sum}--\eqref{cons:box}, we construct the Lagrangian function

\begin{equation}
\begin{aligned}
&\mathcal{L(\mathbf{P},\nu,\boldsymbol{\mu},\boldsymbol{\omega})} = \sum_{k=1}^{K} \frac{c_k}{\ln(1 + a_k P_{AP,k})}\\
& + \nu\!\left(\sum_{k=1}^{K} P_{AP,k}
 - P_{\text{total}}\right)
- \sum_{k=1}^{K} \mu_k P_{AP,k}\\
& + \sum_{k=1}^{K} \omega_k (P_{AP,k}-P_{\max}),
\end{aligned}
\end{equation}
where $\nu \ge 0$, $\mu_k \ge 0$, and $\omega_k \ge 0$ are the dual variables. The KKT conditions are
\begin{equation}
\begin{aligned}
&\frac{\partial \mathcal{L}}{\partial P_{AP,k}} \;=\;
-\frac{c_k a_k}{(1+a_k P_{AP,k})\,[\ln(1+a_k P_{AP,k})]^2}\\
&+\nu - \mu_k + \omega_k \;=\;0,\quad \forall k; \label{kkt:stat}\\[2mm]
&(48)(49);\\
&\nu \ge 0,\ \mu_k \ge 0,\ \omega_k \ge 0;\\
&\nu\!\left(\sum_k P_{AP,k}-P_{\text{total}}\right)=0,\;\\
&\mu_k P_{AP,k}=0,\;
\omega_k (P_{AP,k}-P_{\max})=0,\ \forall k.
\end{aligned}
\end{equation}

If user $k$ is at the optimal point where $0 < P_{AP,k} < P_{\max}$, then by complementary slackness, we have $\mu_k = \omega_k = 0$. From \eqref{kkt:stat}, we get:
\begin{equation}
\nu \;=\; \frac{c_k a_k}{(1+a_k P_{AP,k})\,[\ln(1+a_k P_{AP,k})]^2}.
\label{eq:nu-eq}
\end{equation}
Let $x_k \triangleq 1 + a_k P_{AP,k} > 1$ and $t_k \triangleq \ln x_k > 0$, then:
\begin{equation}
\nu \;=\; \frac{c_k a_k}{x_k t_k^2}
\;\;\Longleftrightarrow\;\;
t_k^2 e^{t_k} \;=\; \frac{c_k a_k}{\nu}.
\end{equation}
Note that $t_k^2 e^{t_k} = \left(t_k e^{t_k/2}\right)^2$, taking the positive root ($t_k > 0$) yields:
\begin{equation}
t_k e^{t_k/2} \;=\; {\sqrt{c_k a_k/\nu}}.
\end{equation}
Define $u_k \triangleq \dfrac{t_k}{2}$, so:
\begin{equation}
u_k e^{u_k} \;=\; \frac{\sqrt{c_k a_k/\nu}}{2}.
\end{equation}
Using the definition of the Lambert W function, $W(z)e^{W(z)} = z$, we have
\begin{equation}
u_k \;=\; W\!\left(\frac{\sqrt{c_k a_k/\nu}}{2}\right), 
\qquad 
t_k \;=\; 2W\!\left(\frac{\sqrt{c_k a_k/\nu}}{2}\right).
\end{equation}
Thus
\begin{equation}
x_k \;=\; e^{t_k} \;=\; e^{2W(z_k)} 
= \left(\frac{z_k}{W(z_k)}\right)^2,
\quad 
z_k \triangleq \frac{\sqrt{c_k a_k/\nu}}{2},
\end{equation}
and the closed-form power solution at interior points is:
\begin{equation}\label{eq:P-int}
\begin{aligned}
P^{\text{int}}_{AP,k}(\nu) 
&\;=\; \frac{x_k-1}{a_k}
\;=\; \frac{1}{a_k}\left[\left(\frac{z_k}{W(z_k)}\right)^2 - 1\right],\\
&z_k \triangleq \frac{\sqrt{c_k a_k}{\nu}}{2}.
\end{aligned}
\end{equation}

Considering the box constraints, the optimal power is given by
\begin{equation}\label{eq:P-proj}
\begin{aligned}
&P_{AP,k}^{\star}(\nu) \;=\; 
\left[\, P^{\text{int}}_{AP,k}(\nu)\, \right]_{0}^{P_{\max}}
\;\triangleq\;\\
&\min\Big\{\,\max\{\,P^{\text{int}}_{AP,k}(\nu),\,0\,\},\; P_{\max}\Big\}.
\end{aligned}
\end{equation}
It can be shown that $P^{\text{int}}_{AP,k}(\nu)$ is monotonically decreasing with respect to $\nu$. Therefore, a bisection method can be used to solve for the dual variable $\nu^\star$, such that:
\begin{equation}
\sum_{k=1}^K P_{AP,k}^{\star}(\nu^\star) \;=\; \min\!\left\{\,P_{\text{total}},\; \sum_{k=1}^K P_{\max}\right\}.
\end{equation}
After solving for $\nu^\star$, we obtain the global optimal solution: $P_{AP,k}^{\star} = P_{AP,k}^{\star}(\nu^\star)$.

Summarizing \eqref{eq:P-int}--\eqref{eq:P-proj}, the final optimal downlink power allocation is:
\begin{equation}
\begin{aligned}
&P_{AP,k}^{\star} \;=\; 
\left[
\frac{1}{a_k}\left(\left(\frac{z_k}{W(z_k)}\right)^2 - 1\right)
\right]_{0}^{P_{\max}},\\
&z_k \;=\; \frac{1}{2}\sqrt{\frac{c_k a_k}{\nu^\star}},
\quad
c_k=\frac{D_{\text{DL},k}\ln 2}{B},\;
a_k=\frac{|h_{\text{eff},k}|^2}{\sigma^2}
\;    
\end{aligned}
\end{equation}
where $W(\cdot)$ is the Lambert W function, and $\nu^\star$ is found using bisection to satisfy $\sum_k P_{AP,k}^{\star} = P_{\text{total}}$. If some users saturate (i.e., $P_{AP,k}^{\star}=0$ or $P_{\max}$), the interval projection automatically handles this.
\paragraph{Remarks}
(1) If no per-user power limit exists ($P_{\max} \to \infty$), the interval projection can be omitted from the above formula; 
(2) If a minimum downlink rate per user constraint $R_{\text{DL},k} \ge R_{\min,k}$ is added, the constraint $P_{AP,k} \ge \frac{1}{a_k}\left(2^{R_{\min,k}/B}-1\right)$ can be included in \eqref{cons:box}, and the interval projection should be adjusted accordingly.

\section{Alternating Optimization Algorithm and the Complexity Analysis}

To solve the original formulated optimization  problem, we adopt an alternating optimization strategy. In each iteration, the strategy optimizes one variable group while keeping the others fixed. This iterative approach continues until the objective converges or a maximum number of iterations is reached.

\vspace{1em}
\subsection{Overall Alternating Optimization Algorithm}
The overall optimization procedure follows an alternating update logic, where each variable block is iteratively refined while keeping the others fixed. This process can be summarized in the following pseudocode for the proposed alternating optimization algorithm:
\begin{algorithm}[H]
\caption{Proposed Alternating Optimization Scheme}
\begin{algorithmic}[1]
\STATE \textbf{Initialize:} Randomly initialize $\{m_k^{(0)}\}$, $\{\alpha_k^{(0)}\}$, $\{\Delta f_k^{(0)}\}$, $\boldsymbol{\Theta}^{(0)}$, and $\{P_k^{\mathrm{AP},(0)}\}$.
\STATE Set iteration index $t = 0$ and maximum iteration $T_{\max}$.
\REPEAT
    \STATE $t \leftarrow t + 1$
    \STATE \textbf{Step 1:} Update model split layers $\{m_k^{(t)}\}$ using a exhaustive search
    \STATE \textbf{Step 2:} Update DT selection variables $\{\alpha_k^{(t)}\}$ based on local vs. virtual execution delay.
    \STATE \textbf{Step 3:} Optimize DT frequency offsets $\{\Delta f_k^{(t)}\}$ via Lagrangian and KKT conditions.
    \STATE \textbf{Step 4:} Optimize IRS phase matrix $\boldsymbol{\Theta}^{(t)}$ via SDR-based method.
    \STATE \textbf{Step 5:} Optimize AP downlink power allocation $\{P_k^{\mathrm{AP},(t)}\}$ using convex optimization.
\UNTIL{convergence of objective or $t \geq T_{\max}$}
\RETURN Final solution $\{m_k\}, \{\alpha_k\}, \{\Delta f_k\}, \boldsymbol{\Theta}, \{P_k^{\mathrm{AP}}\}$
\end{algorithmic}
\end{algorithm}

\subsection{Complexity Analysis}

We now analyze the computational complexity of each step in a single iteration:

\begin{itemize}
    \item \textbf{Split Layer Optimization (\(m_k\)):} For each UD, this involves evaluating $M$ choices. Assuming $K$ UDs and constant cost per evaluation (forward pass), the complexity is \(\mathcal{O}(KM)\)\cite{8279531,10924408,8700264}.
    
    \item \textbf{DT Assignment (\(\alpha_k\)):} For each UD, compute and compare local vs. DT delay, leading to complexity \(\mathcal{O}(K)\).
    
    \item \textbf{DT Frequency Offset (\(\Delta f_k\)):} Each UD’s offset is solved via a closed-form or lightweight iterative method from KKT conditions. The overall complexity is \(\mathcal{O}(K)\)\cite{10054442,6832526}.
    
    \item \textbf{IRS Phase Matrix Optimization (\(\boldsymbol{\Theta}\)):} This is solved using SDR, which has polynomial complexity in the number of IRS elements $N_{\mathrm{IRS}}$. Specifically, solving a SDP of dimension $N_{\mathrm{IRS}}$ incurs complexity \(\mathcal{O}(N_{\mathrm{IRS}}^6)\) in the worst case, though often lower in practice\cite{8485724,10290939}.
    
    \item \textbf{Downlink Power Allocation (\(P_k^{\mathrm{AP}}\)):} This convex optimization over $K$ variables is solvable in polynomial time, with interior-point methods yielding complexity \(\mathcal{O}(K^3)\).
\end{itemize}
Let $T$ denote the number of outer iterations. The total computational complexity is
\begin{equation}
    \mathcal{O}\left(T \cdot \left(KM + K + K + N_{\mathrm{IRS}}^6 + K^3 \right)\right),
\end{equation}
which simplifies to
\begin{equation}
    \mathcal{O}\left(T \cdot \left(KM + N_{\mathrm{IRS}}^6 + K^3 \right)\right).
\end{equation}

\subsection{Convergence Analysis}

The proposed alternating optimization algorithm follows the principle of block coordinate descent (BCD). 
At each iteration $t$, one block of variables 
$\{m_k\}$, $\{\alpha_k\}$, $\{\Delta f_k\}$, $\Theta$, or $\{P_{\text{AP},k}\}$ 
is updated while keeping the others fixed. 
Since each subproblem is solved either optimally in closed-form (e.g., $\Delta f_k$ update) 
or via convex optimization (e.g., IRS phase optimization through SDR, power allocation via convex programming), 
the overall objective function is guaranteed to be non-increasing, which can be expressed as 

\begin{equation}
J^{(t+1)} \leq J^{(t)}, \quad \forall t,
\end{equation}

where $J^{(t)}$ denotes the objective value at the $t$-th iteration. 

Moreover, the total objective function is defined as the weighted sum of latency and training loss:
\begin{equation}
J = \sum_{k=1}^K T^{\text{total}}_k + \lambda \cdot \mathcal{L}_{\text{train}}, 
\end{equation}
which is lower-bounded by zero, i.e.,
\begin{equation}
J^{(t)} \geq 0, \quad \forall t.
\end{equation}

Therefore, the sequence $\{J^{(t)}\}$ generated by the algorithm is monotonically non-increasing and bounded from below, 
which guarantees convergence to a finite value:
\begin{equation}
\lim_{t \to \infty} J^{(t)} = J^\star.
\end{equation}

However, due to the presence of integer decision variables $m_k$ and $\alpha_k$, 
the algorithm cannot guarantee convergence to the global optimum. 
Instead, it converges to a stationary solution (local optimum), 
which is sufficient for practical deployment, as also validated in the simulation results.

\section{Simulation Results}

In our simulations, we consider a wireless edge inference system consisting of a multi-antenna AP, a single IRS, and $K=100$ single-antenna UDs. The AP is equipped with $N_{\text{AP}} = 32$ antennas, and the IRS comprises $N_{\text{IRS}} = 32$ passive reflecting elements. The direct link between the AP and UDs is assumed to be severely blocked and thus ignored. All communication relies on the cascaded link through the IRS.

We adopt a quasi-static flat-fading channel model where the channels remain constant within each transmission frame. Both the channel of link AP-to-IRS $\mathbf{G} \in \mathbb{C}^{N_{\text{AP}} \times N_{\text{IRS}}}$ and the channel of link IRS-to-UD $k$ $\mathbf{h}_k^{(\text{IRS})} \in \mathbb{C}^{N_{\text{IRS}} \times 1}$ are modeled as independent Rayleigh fading channels with small-scale fading coefficients drawn from $\mathcal{CN}(0,1)$. Large-scale path loss is incorporated using the model:
\begin{equation}
\mathrm{PL}(d) = C_0 \left( \frac{d}{d_0} \right)^{-\alpha},
\end{equation}
where $C_0$ denotes the path loss at the reference distance $d_0 = 1$ meter (set to $-30$ dB in linear scale), $d$ is the link distance, and $\alpha$ is the path loss exponent. We set $\alpha = 2.2$ for AP-to-IRS links and $\alpha = 2.8$ for IRS-to-UD links, capturing typical urban macro-cell scenarios. The effective channel for each UD is computed as the product of the AP-to-IRS and IRS-to-UD channels with optimized IRS phase shifts.

Unless otherwise specified, the total downlink transmit power at the AP is set to $P_{\text{total}} = 3000$ mW. The noise power at each UD is fixed at $\sigma^2 = 10^{-11}$ W. The required minimum data rate per UD is set to $R_{\text{target}} = 5$ kbps. All neural network models used for edge inference are assumed to have $M = 12$ layers, with the layer-wise computational complexity $C_{k,m}$ for UD $k$ and layer $m$ randomly drawn from a uniform distribution between $0.5$ and $1.5$ GFLOPs. Local device computation frequencies $f_k^{(\text{local})}$ are drawn from a uniform distribution between $5$ and $12$ GHz. The energy-efficiency parameter for dynamic frequency scaling is set to $\kappa = 10^{-28}$\cite{HuangTWC2019,WuJSAC2023,YouTWC2017,8488502,10124759,9685493}.

We evaluate different baseline strategies including full local inference, full offloading via IRS, a genetic algorithm GA-based hybrid split strategy, and an ADMM-based method, under various system configurations and power constraints.
\subsection{Baseline Strategies}

To benchmark the performance of our proposed method, we evaluate the following baseline strategies:

\textbf{1) Full Local Inference (No IRS):} Each UD performs the entire inference process locally without any offloading. This scheme reflects the traditional device-only computation paradigm. The total delay for each UD is determined by the local processing time, which depends on the accumulated computational complexity of all layers and the UD’s local CPU frequency. IRS and AP resources are not utilized in this scheme.

\textbf{2) Full Offloading via IRS (No DT):} In this strategy, all inference tasks are fully offloaded to the AP via the IRS. Each UD transmits all intermediate features to the AP after passing through the entire network. Downlink power allocation is optimized to ensure a minimum required rate per UD, subject to a total transmit power constraint. Since UDs are unable to perform any computation locally, the system must rely entirely on wireless transmission and AP-side computation. Dynamic frequency scaling (DT) is disabled in this case.

\textbf{3) GA-based Hybrid Inference:} This approach employs a Genetic Algorithm (GA) to determine the optimal layer-wise inference split point for each UD\cite{8944822}. The objective is to minimize the weighted sum of communication delay, local computation delay, and inference loss. For each UD, a split layer is selected such that partial inference is conducted locally, and the remaining layers are offloaded to the AP via IRS. The AP downlink power allocation is optimized accordingly under a total power constraint. Dynamic frequency scaling is enabled to improve energy efficiency at the UD side when offloading is required.

\textbf{4) ADMM-based Hybrid Inference:} Alternating Direction Method of Multipliers (ADMM) is utilized to solve the same split inference problem in a distributed optimization framework\cite{9461690,8089407}. In each iteration, the local model is trained on-device, followed by an update of the user-specific split point through an ADMM-based optimization subroutine. The framework balances local computation and wireless transmission, with dynamic frequency adjustment applied when beneficial. This approach promotes convergence toward an efficient hybrid inference configuration.

\subsection{Numerical Analyses}

\begin{figure}[t]
    \centering
    \includegraphics[width=0.48\textwidth]{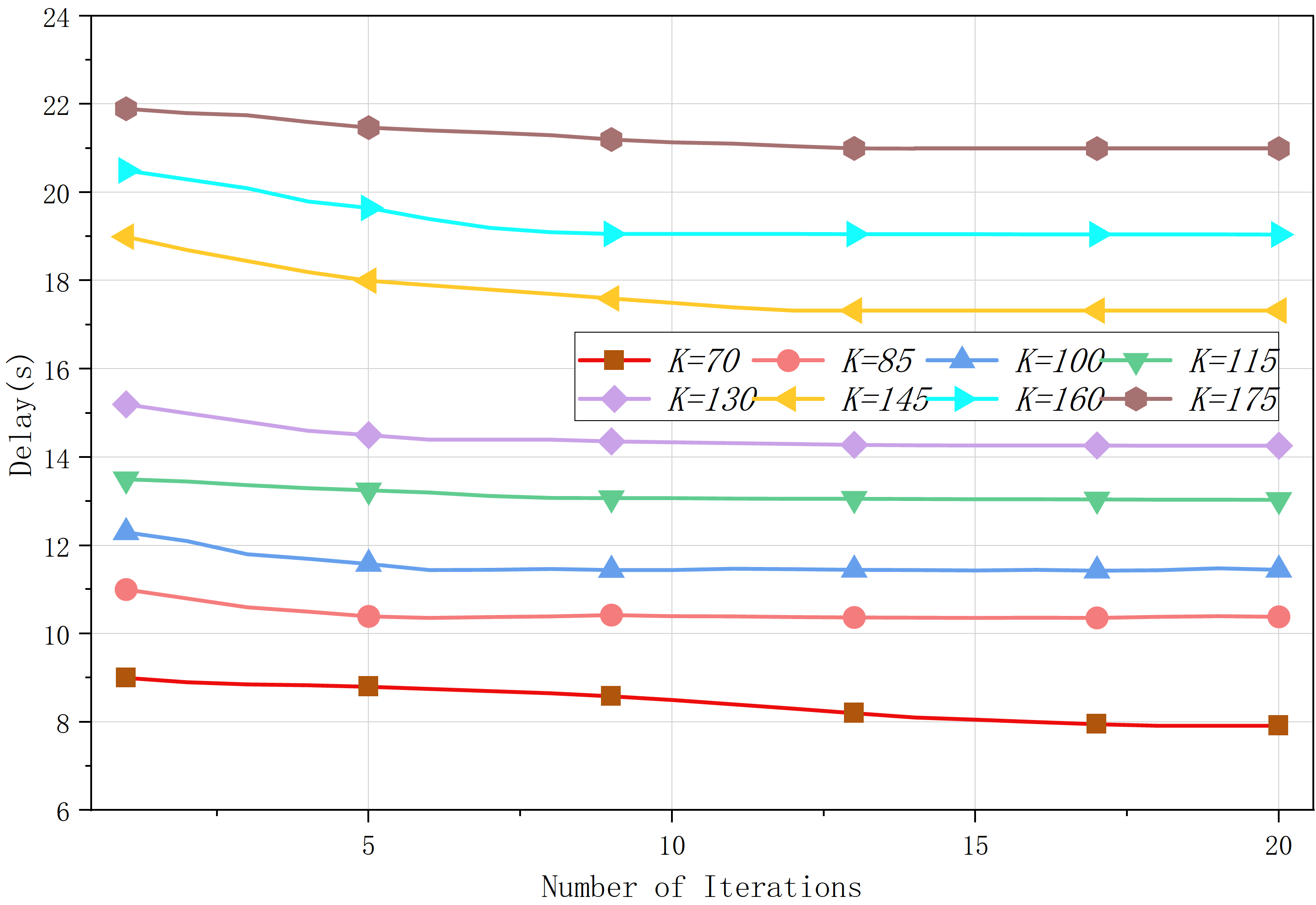}
    \caption{Convergence of delay performance under different numbers of UDs $K$.}
    \label{fig:delay_vs_}
\end{figure}

Fig.~\ref{fig:delay_vs_} illustrates the variation of system delay with respect to the number of optimization iterations for different numbers of UDs $K$. It can be observed that, for all $K$ values, the delay decreases gradually as the number of iterations increases, indicating the effectiveness of the iterative optimization process in improving the delay performance. The improvement is more significant during the first few iterations and gradually saturates as the algorithm converges.
In addition, the absolute delay level is strongly dependent on the number of UDs $K$. When $K$ is small (e.g., $K=70$), the system delay is the lowest due to the reduced computational and communication load. As $K$ increases, the delay increases almost monotonically, which can be attributed to two main reasons: 
1) a larger $K$ introduces higher communication overhead since more UDs share the same spectral and power resources; 
2) the aggregated computational demand grows with $K$, resulting in increased local processing time and queuing delays at the AP. 
For large $K$ values (e.g., $K=160$ and $K=175$), the delay remains at a relatively high level even after convergence, showing the limitation of available resources under heavy network load.
Overall, this figure demonstrates that the proposed iterative scheme can effectively reduce the delay for various $K$ values, but the achievable performance is bounded by the inherent trade-off between resource allocation and the number of active UDs in the system.

\begin{figure}[htbp]
    \centering
    \subfloat[$K=55$]{
        \includegraphics[width=0.45\textwidth]{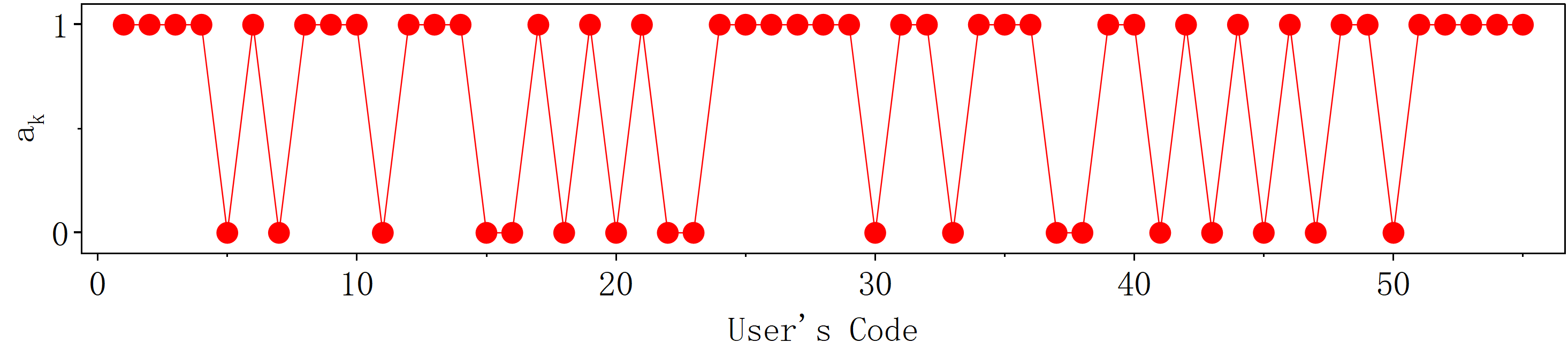}
        }\\
    \subfloat[$K=85$]{
        \includegraphics[width=0.45\textwidth]{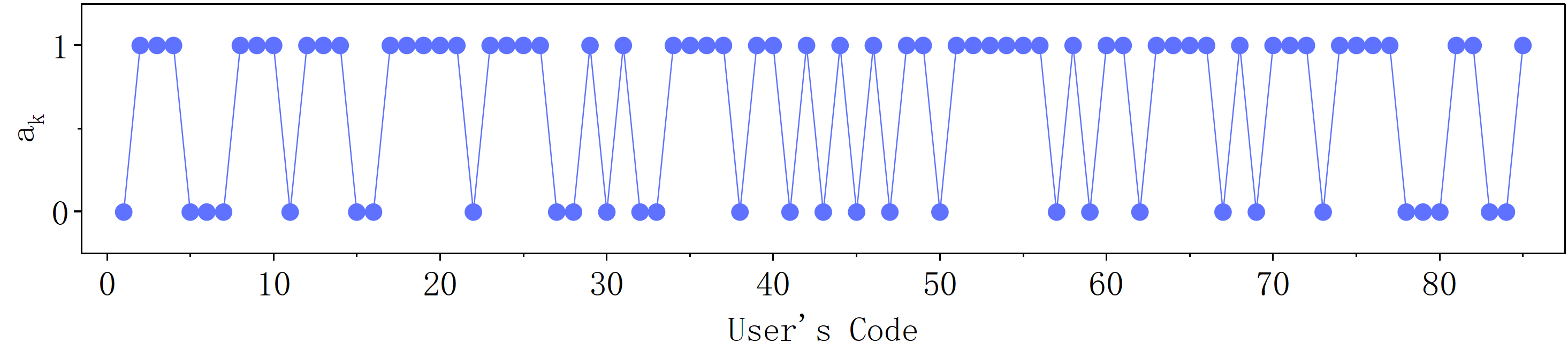}
        }\\
    \subfloat[$K=115$]{
        \includegraphics[width=0.45\textwidth]{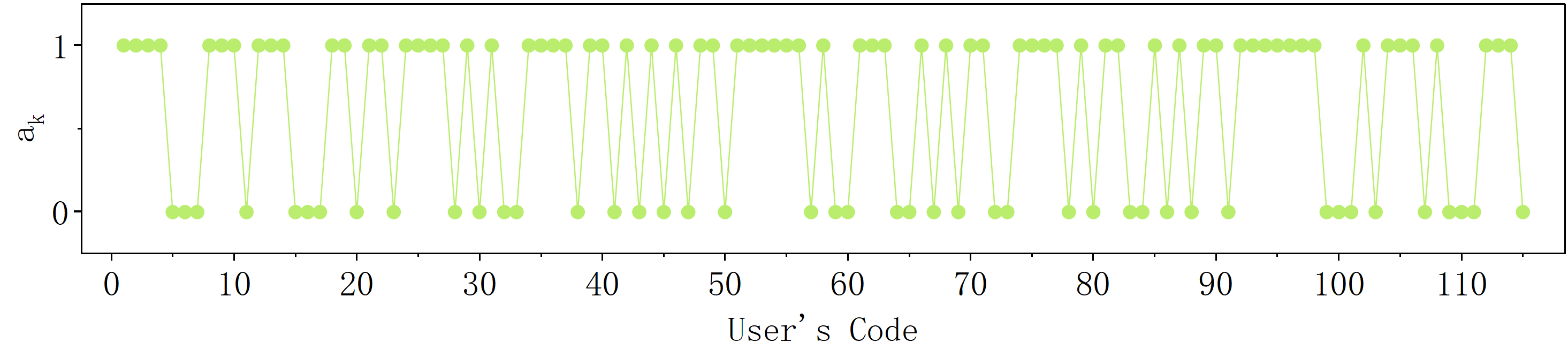}
        }\\
    \subfloat[$K=145$]{
        \includegraphics[width=0.45\textwidth]{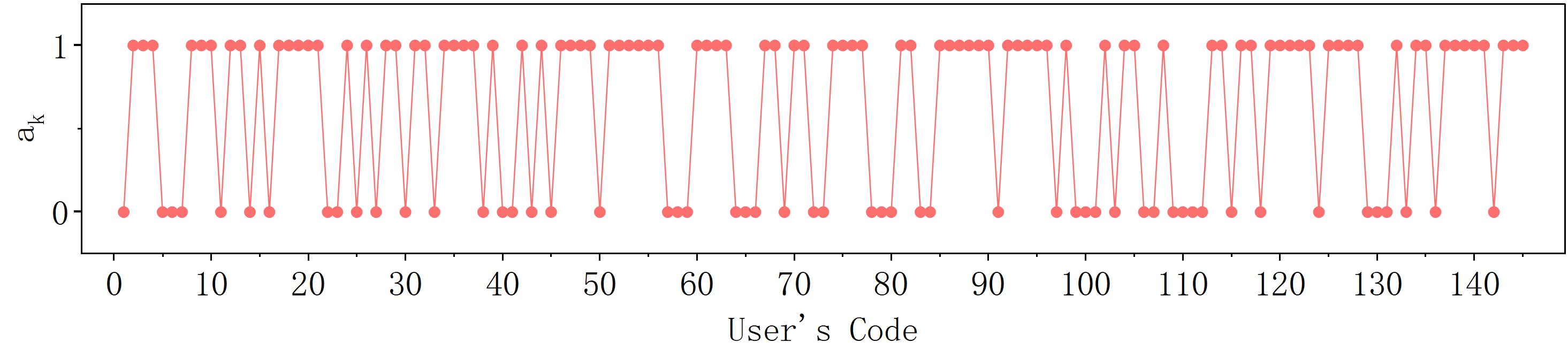}
        }\\
    \subfloat[$K=175$]{
        \includegraphics[width=0.45\textwidth]{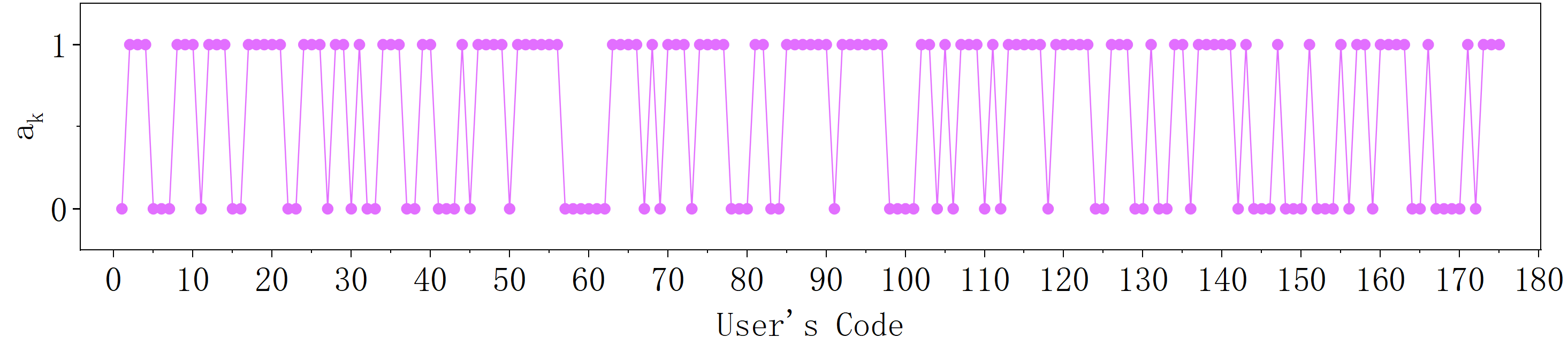}
        }\\
    \caption{Binary decision variable $\alpha_k$ for different numbers of UDs. $\alpha_k = 1$ indicates local computation, while $\alpha_k = 0$ indicates computation offloading to the DT. As $K$ increases, more UDs are assigned to DT due to limited local computing resources and stricter delay constraints.}
    \label{fig:ak_distribution}
\end{figure}

\par Fig.~\ref{fig:ak_distribution} illustrates the binary decision variable $\alpha_k$ for different numbers of UDs, where $\alpha_k = 1$ indicates that the computation task of UD $k$ is executed locally, while $\alpha_k = 0$ implies that the task is offloaded to the digital twin (DT) at the server. The subfigures correspond to $K = 55, 85, 115, 145,$ and $175$, respectively. It can be observed that as $K$ increases, the distribution of $\alpha_k$ becomes more irregular, with a larger proportion of UDs being offloaded to DT. This is due to the limited local computation capacity and stricter delay requirements, which make DT processing more favorable for UDs with insufficient computing power. These patterns also reflect the heterogeneity in UDs' computational capabilities and the task complexity distribution.

\begin{figure}[htbp]
    \centering
    \includegraphics[width=0.45\textwidth]{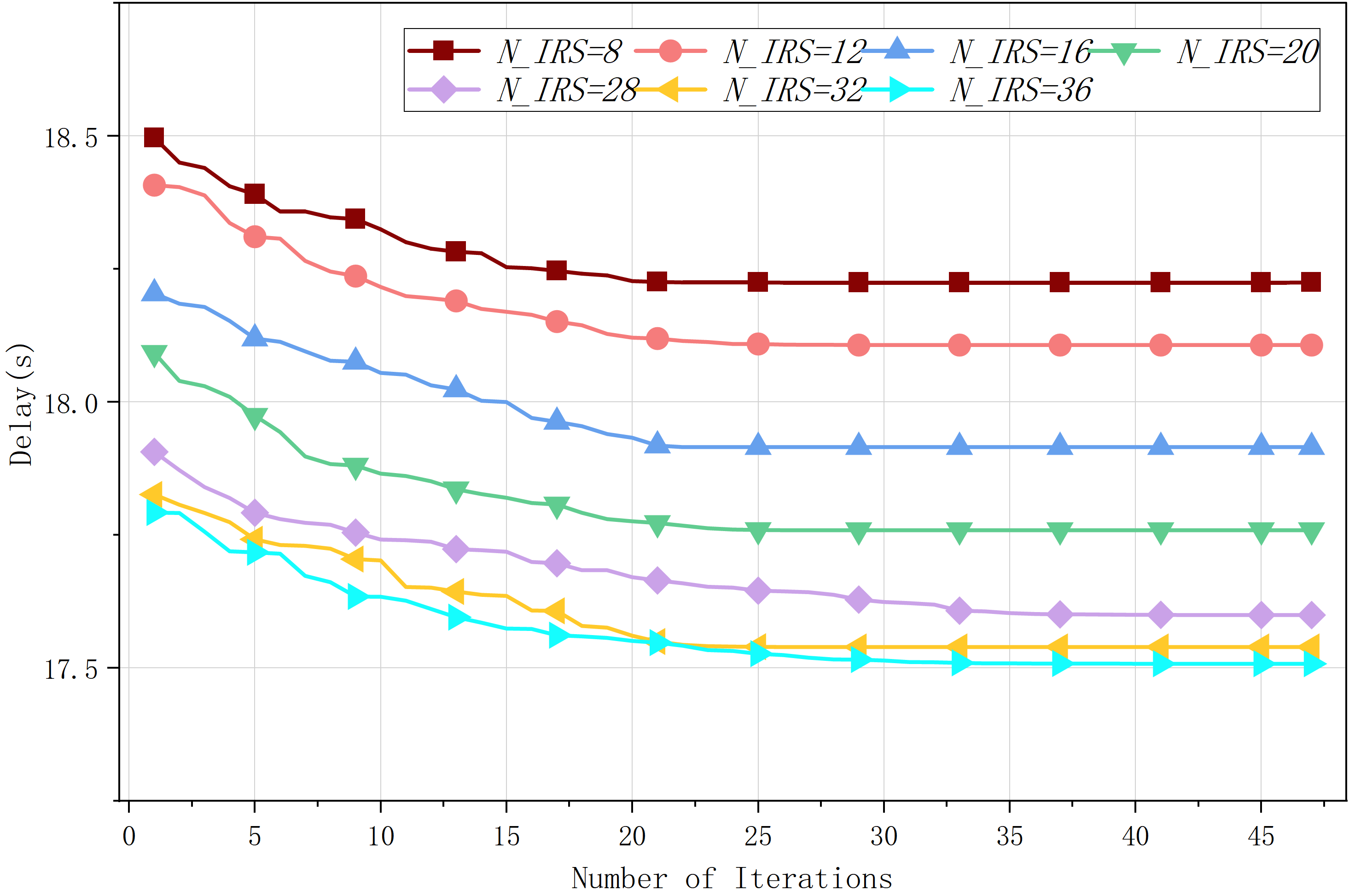}
    \caption{Convergence of system delay under different numbers of IRS reflecting elements $N_{\text{IRS}}$.}
    \label{fig:delay_vs_NIRS}
\end{figure}

\par Fig.~\ref{fig:delay_vs_NIRS} illustrates the variation of the total system delay with the number of iterations for different values of IRS reflecting elements $N_{\text{IRS}}$. It can be observed that, for all considered $N_{\text{IRS}}$ values, the delay decreases rapidly during the initial iterations and then gradually converges. A larger $N_{\text{IRS}}$ consistently leads to a lower final delay, owing to the enhanced beamforming gain and improved channel quality brought by more reflecting elements. For example, when $N_{\text{IRS}}=36$, the system achieves the lowest delay due to the higher spatial degrees of freedom, which facilitate better signal alignment and reduce transmission time. Conversely, smaller $N_{\text{IRS}}$ values (e.g., $N_{\text{IRS}}=8$) exhibit higher delays because of the limited capability in improving the effective channel gain. This trend highlights the importance of IRS deployment size in achieving superior communication-computation performance in the proposed system.

\begin{figure}[htbp]
    \centering
    \includegraphics[width=0.45\textwidth]{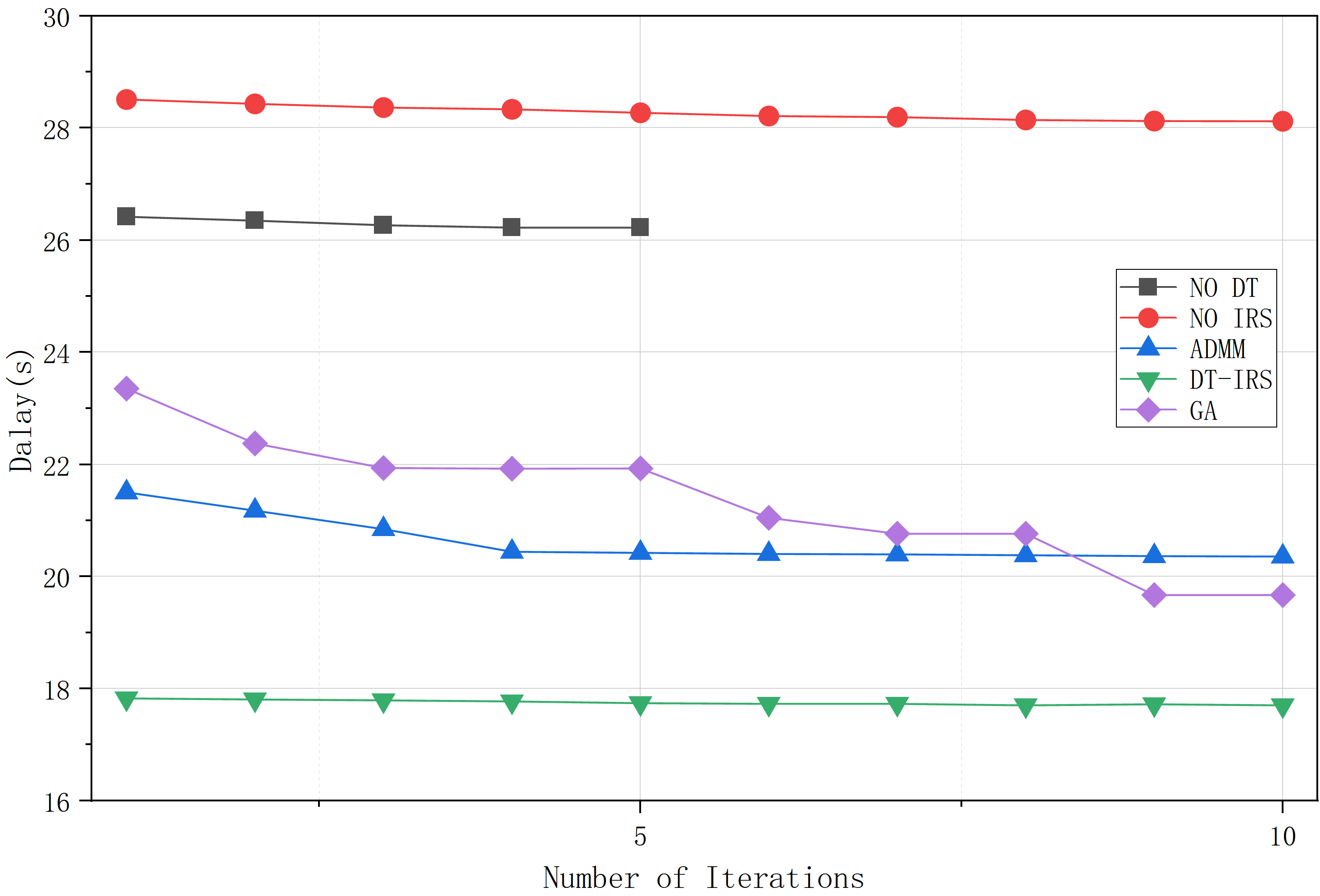}
    \caption{Comparison of convergence performance in terms of delay under different baseline schemes.}
    \label{fig:baseline_delay}
\end{figure}

\par Fig.~\ref{fig:baseline_delay} presents the system delay performance for different baseline schemes, including the proposed DT-IRS method, No IRS, No DT, ADMM-based optimization, and GA-based optimization. It is evident that the proposed DT-IRS approach achieves the lowest delay across all iterations, converging quickly to approximately $17.8$~s, owing to the joint optimization of IRS phase shifts, and DT-assisted computation, which maximizes both communication and computation efficiency. The No IRS and No DT baselines exhibit the highest delays (above $26$~s and $28$~s, respectively), indicating the significant performance degradation when either IRS-assisted channel enhancement or DT-based computation offloading is absent. The ADMM-based method achieves better performance than No IRS/No DT but remains inferior to the proposed approach due to its reliance on a convexified optimization framework, which limits the achievable solution space. The GA-based method initially shows higher delay than ADMM but eventually converges to a slightly lower value, benefiting from its global search capability, though at the cost of slower convergence. Overall, the results highlight that the synergy between IRS and DT is critical to minimizing delay, and that advanced joint optimization significantly outperforms simplified or component-removed baselines.

\begin{figure}[htbp]
    \centering
    \includegraphics[width=0.45\textwidth]{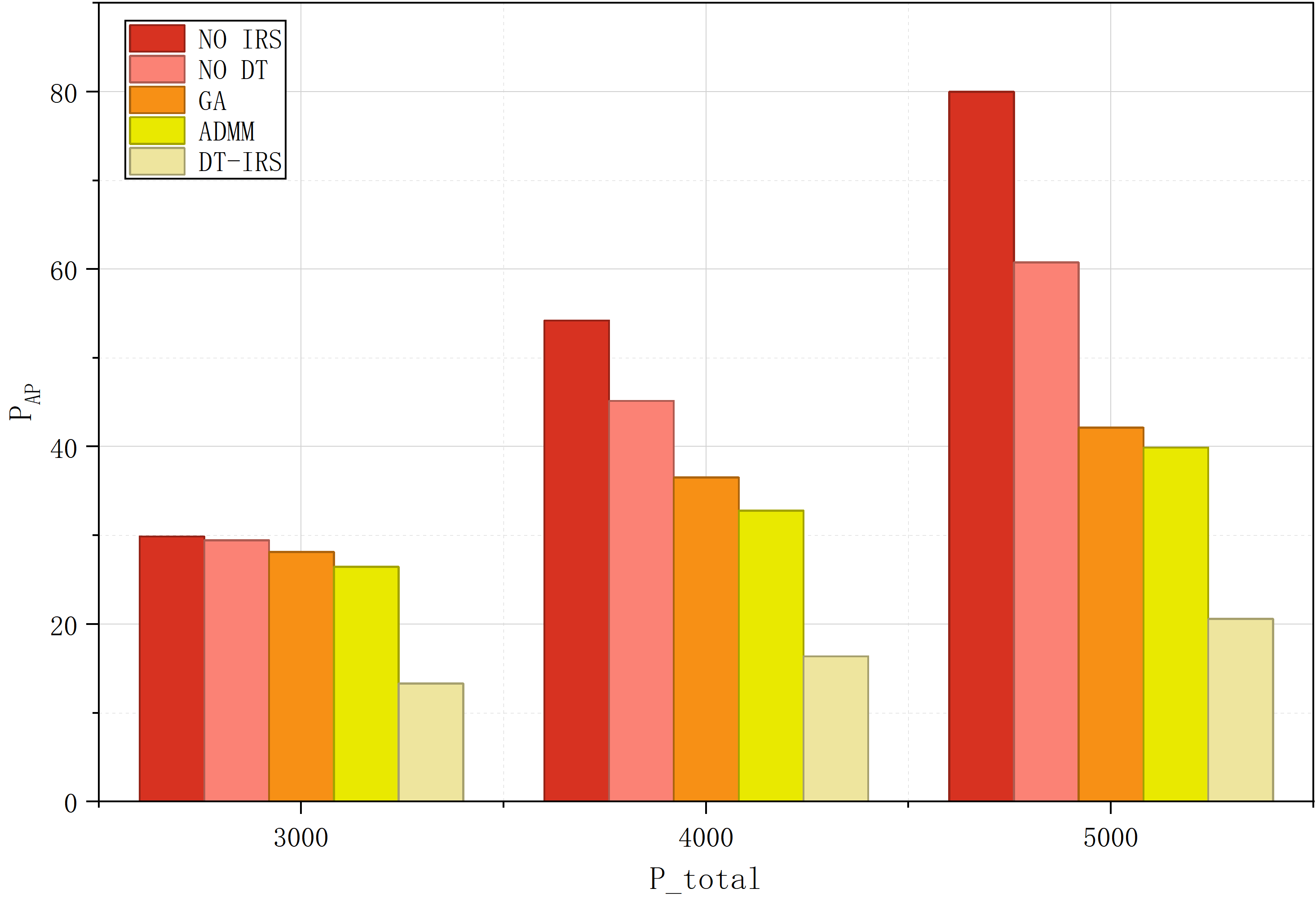}
    \caption{Average transmit power at the AP ($P_{\mathrm{AP}}$) under different $P_{\mathrm{total}}$ settings for various schemes.}
    \label{fig:pap_comparison}
\end{figure}

\par Figure~\ref{fig:pap_comparison} illustrates the average AP transmit power consumption $P_{\mathrm{AP}}$ for five schemes---Without IRS, Without DT, GA-based optimization, ADMM-based optimization, and the proposed DT-IRS method---under different maximum total power budgets $P_{\mathrm{total}} \in \{3000, 4000, 5000\}$~mW. Across all $P_{\mathrm{total}}$ values, the proposed DT-IRS approach consistently yields the lowest $P_{\mathrm{AP}}$, indicating that the joint optimization of IRS phase shifts, DT-assisted computation, and power allocation allows the system to achieve target performance with significantly reduced transmission power. In contrast, the Without IRS and Without DT baselines exhibit substantially higher $P_{\mathrm{AP}}$, with power demand increasing steeply as $P_{\mathrm{total}}$ grows; this is due to the lack of either channel enhancement or computational offloading, which forces the AP to transmit at higher power to meet communication and computation deadlines. The GA-based and ADMM-based methods perform better than the two baselines but still require more transmit power than the proposed scheme, as their optimization frameworks either lack full joint variable coupling (i.e., ADMM) or incur suboptimal resource allocation in early iterations (i.e., GA). Notably, the gap between DT-IRS and other methods widens with larger $P_{\mathrm{total}}$, demonstrating that the proposed method more efficiently exploits additional power resources to further reduce AP transmit requirements.

\begin{figure}[htbp]
    \centering
    \includegraphics[width=0.45\textwidth]{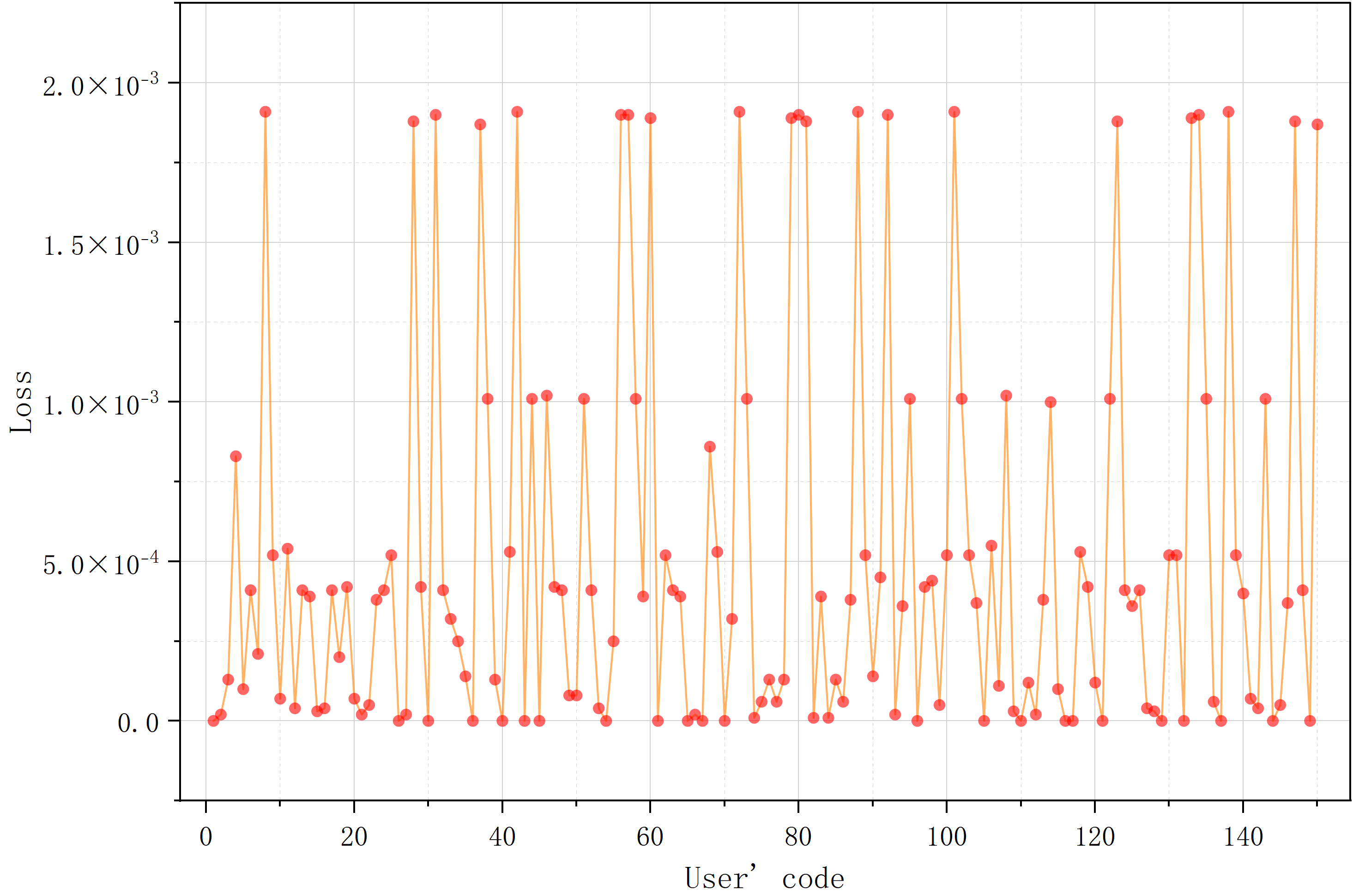}
    \caption{Variation of training loss with different numbers of UDs $K$.}
    \label{fig:loss_vs_K}
\end{figure}

Fig.~\ref{fig:loss_vs_K} depicts the variation in the model training loss across different values of $K$. The loss remains at a relatively low magnitude (on the order of $10^{-3}$) for all tested $K$, with small fluctuations arising from the inherent randomness of data generation and network initialization. Although occasional spikes are observed, these are not persistent and quickly return to the baseline level, indicating that the overall training process is stable for varying UD numbers. This suggests that the model maintains robustness in convergence behaviour regardless of the scale of participating UDs.

\section{Conclusion}
In this paper, we have proposed an integrated edge intelligence framework that jointly optimizes communication and computation for large-model split learning, leveraging IRS for channel enhancement and DT technology as a computation backup mechanism. We have formulated the joint optimization problem and developed an alternating optimization algorithm to solve it effectively. The optimized solution dynamically determines the optimal split point of the deep neural network based on network conditions and computational resources, while adaptively configuring IRS phase shifts, AP transmit power, and DT processing frequency to minimize the overall delay. Extensive simulation results have  demonstrated that the proposed framework significantly reduces total delay compared to baseline methods, while maintaining model accuracy and ensuring robust performance under dynamic network states.

\textbf{Future Work:} Our future research will focus on extending the proposed framework to support more heterogeneous devices and diverse model architectures, as well as incorporating real-world deployment constraints such as asynchronous updates and energy harvesting capabilities. Additionally, we plan to investigate learning-based control policies that enable faster adaptation to time-varying network and computation conditions. Although this work considers a single AP for clarity, the proposed framework can be readily extended to multiple cooperative APs. In such a scenario, the uplink transmissions of intermediate features can be jointly received and decoded by multiple APs, while downlink gradients or inference results can be broadcast via coordinated beamforming. The DT replicas may be instantiated at different APs, and their computing frequencies dynamically adjusted to balance the load across the network. The optimization problem would then be extended to include AP–user association and inter-AP coordination, which can be addressed by distributed or hierarchical alternating optimization methods. We leave this as a promising direction for future work.

\bibliographystyle{IEEEtran} 
\bibliography{IEEEabrv,ref} 

\vspace{11pt}


\vfill

\end{document}